\documentclass{article}

\usepackage{PRIMEarxiv}

\usepackage[utf8]{inputenc} 
\usepackage[T1]{fontenc}    
\usepackage{hyperref}       
\usepackage{url}            
\usepackage{booktabs}       
\usepackage{amsfonts}       
\usepackage{nicefrac}       
\usepackage{microtype}      
\usepackage{lipsum}
\usepackage{fancyhdr}       
\usepackage{graphicx}       
\graphicspath{{media/}}     
\usepackage{wrapfig}
\usepackage{color}
\usepackage{float}
\usepackage{listings}
\usepackage{amsmath}
\usepackage{graphicx}
\usepackage{makecell}
\usepackage{tabularx}
\usepackage{setspace}
\title{Investigating Skin Temperature-Based Overheating in mmWave Smartphones: Power and Thermal Models for Optimal Non-Throttling Performance
}

\author{
  Henglin Pu, Xingqi Wu \\
  University of Michigan, Ann Arbor\\
  \texttt{\{henglinp, xingqiwu\}@umich.edu} \\
}

\begin{document}
\maketitle

\begin{abstract}

5G mmWave, as a revolutionary cellular technology, holds monumental potential for innovations in many academic and industrial areas. However, widespread adoption of this technology is hindered by the severe overheating issues experienced by current Commercial Off-The-Shelf (COTS) mmWave smartphones. This study aims to identify the root causes of device skin temperature related throttling during 5G transmission, and to quantify power reduction required to prevent such throttling in a given ambient temperature. The key insight of our paper is leveraging the power model and thermal model of mmWave smartphone to acquire the quantitative relationship among power consumption, ambient temperature and device skin temperature. This approach allows us to determine the extent of power reduction required to prevent throttling under specific ambient temperature conditions.
\end{abstract}


\section{INTRODUCTION}

The advent of 5G millimeter wave (mmWave) technology as a next-generation communication system has facilitated the allocation of high-frequency bands above 24 GHz, thereby providing increased capacity delivery and enhanced management of peak rates. Owing to its advantages of high-throughput and low latency, the mmWave paradigm has garnered considerable research interest in various domains, including satellite communication [17][18], high-resolution video streaming and broadcasting [19], augmented and virtual reality (AR/VR) [20], and automotive applications [21].

However, the potential of 5G mmWave technology is accompanied by several challenges that warrant the attention of researchers. One prominent challenge is the temperature-related throttling problem, which can significantly impede 5G mmWave throughput. Recent literature [1] has demonstrated that the skin temperature of mobile devices markedly influences 5G mmWave throughput. The study reveals that mmWave throughput undergoes throttling and becomes restricted when the skin temperature reaches a certain threshold. As 5G mmWave exhibits a considerably faster transmission rate compared with Long-Term Evolution (LTE) and 5G mid-bands, it leads to an increased workload for both the smartphone central processing unit (CPU) and transceiver. To effectively address the overheating issue, it is essential to comprehend its underlying cause, i.e., whether the CPU or transceiver contributes to the significant rise in skin temperature.

This paper aims to identify the specific root causes of skin temperature-related throttling during 5G transmission and to determine the extent of power reduction necessary to avoid throttling. To accomplish this, the power consumption of a smartphone during 5G mmWave transmission is examined, and a system-level power model is established. Subsequently, the power model is expanded into a thermal model, which enables the estimation of the relationship between ambient temperature and the power reduction required to prevent throughput throttling, as well as the association between ambient temperature and the maximum achievable throughput without throttling. Based on these models, it is demonstrated that the transceiver is the primary contributor to device skin temperature-based throttling, while the CPU has a relatively minor impact. By pinpointing the root cause of skin temperature-related throttling during 5G transmission and quantifying the necessary power reduction, this paper presents a significant contribution to the advancement of 5G mmWave technology.

Simply put, this paper makes the following contributions:
\begin{itemize}
    \item We build a system-level power model for mmWave smartphone Pixel 5 involving both transceiver and multi-core CPU;
    \item We ascertain that the transceiver serves as the principal factor contributing to the elevation of skin temperature during 5G mmWave transmission, whereas the CPU exhibits only a minimal influence;
    \item We investigate the correlation between the requisite power reduction to avert throughput throttling and ambient temperature through thermal modeling, which holds instructive implications for subsequent optimization.
\end{itemize}

\section{RELATED WORK}

In 2017, the 3rd Generation Partnership Project (3GPP) introduced a procedure termed User Equipment (UE) assistance information [2], which enables UEs to communicate their preferred configurations to the gNodeB (gNB) when experiencing elevated temperatures. As the inaugural study investigating the effects of temperature on mmWave device performance, [3] demonstrates the phenomenon whereby the mmWave antenna temperature increases to an extreme value of 68$^o C$ at a 1.9 Gbps transmission rate, subsequently causing link throughput throttling within a brief period. However, the authors neglect to consider the impact of the central processing unit (CPU) on temperature fluctuations. More recent studies [4][5][10][11] present comprehensive measurements on the power performance of 4G and 5G networks, including mmWave, and develop a power model primarily based on Radio Resource Control (RRC) states. Nonetheless, the influence of the CPU remains unaddressed. In [6], the authors broadly compare energy consumption across various network types and internet service providers (ISPs) in relation to battery levels without presenting any numerical power and energy results.

At present, two primary categories of power-saving approaches exist for 5G New Radio (NR): one involves optimizing Radio Resource Management methods, such as employing Discontinuous Reception (DRX)-related techniques [9][10], introducing RRC inactive states [15][16], and others; the second category focuses on optimizing radio frequency (RF) antenna circuits [12][13][14]. Numerous energy-efficient compression technologies for communication have been proposed [7][8], highlighting the effectiveness of compression in conserving communication power. However, the impact of compression technologies on 5G mmWave power saving has seldom been discussed.

Currently, only a limited number of studies concentrate on the effect of temperature on mmWave devices. None have constructed a power model for mmWave smartphones that considers both the CPU and transceiver, nor have any investigated the influence of the CPU on thermal performance. Furthermore, no existing research has explored the power or energy reduction necessary to prevent throughput throttling, which could hold significant implications for future optimization efforts.

\section{OVERVIEW OF SYSTEM MODEL}

\subsection{Power model}

Given that our primary objective is to determine the power reduction necessary to avert throttling during 5G mmWave transmission, our focus is directed towards the power state featuring RRC connectivity. Consequently, we primarily model the CPU and transceiver without considering other components that are not utilized during 5G mmWave transmission.

\begin{wrapfigure}{r}{0.35\textwidth}
  \begin{center}
    \includegraphics[width=0.25\textwidth]{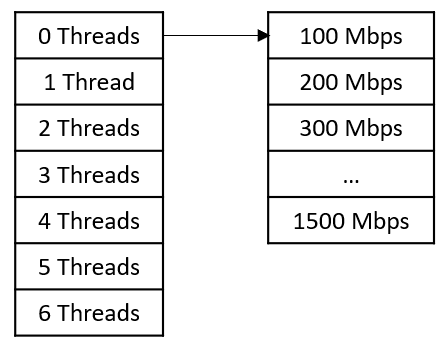}
  \end{center}
  \caption{Data structure for training}
\end{wrapfigure}

Our power model is constructed based on system-level measurements gathered through the application of the linear regression method. This approach facilitates the fitting of total power consumption to a set of system variables with corresponding linear coefficients. Table 1 enumerates the system variables selected for incorporation into the power model.

A crucial step in constructing the model involves training. The primary concept entails using a set of training programs to modify one activity state variable at a time while maintaining all other variables constant. We collect training data for each total CPU usage and downlink throughput separately, under a fixed CPU frequency value. For each total CPU usage, we consider 15 to 19 downlink throughput settings. For each throughput setting, we record the actual total CPU usage value, downlink value, and other system variables under consideration, as well as their corresponding total power consumption within a fixed time interval.

The data structure employed for collecting training data is depicted in Figure 1. Each individual thread pertains to the calculation of the SHA1 value of a 2 MB buffer filled with random values. Initiating one thread will consume 0.125$\%$ of the total CPU usage. We adjust the number of threads to generate a set of total CPU usage values, and we also modify the bandwidth using iPerf3 to produce different throughput values for each total CPU usage setting.

\begin{table}[h]
\centering
\begin{tabular}{|c||c|}
\hline
UE Components Considered & System Variable for Modeling   \\\hline
 CPU & Total CPU Usage(UT), CPU 6 Usage(U6),\\ & CPU 7 Usage(U7)\\\hline
Transciever & Downlink(DL), Uplink(UL), Base Power(BP),\\ & Channel Number(CN) \\\hline
\end{tabular}
\caption{\label{tab:SOTA} All System Variable for Current Model}
\end{table}

Upon gathering power traces for hardware components under the supervision of our training processes, we employ multivariable regression to minimize the sum of squared errors for the power coefficients. The mathematical formulation for the power model, given a specific channel number and CPU frequency, is expressed as follows:

\begin{equation}
\scriptstyle
\left(\begin{array}{c}
P_{0} \\
P_{1} \\
\vdots \\
P_{m}
\end{array}\right)=(1 - I_{5G})\times BP_{CPU} \cdot\left(\begin{array}{c}
1 \\
1 \\
\vdots \\
1
\end{array}\right)+
I_{5G}\times BP_{5G} \cdot\left(\begin{array}{c}
1 \\
1 \\
\vdots \\
1
\end{array}\right) + \left[\begin{array}{ccccc}
UT_{0} & U6_0 & U7_0 & DL_0 & UL_0 \\
UT_{1} & U6_0 & U7_1 & DL_1 & UL_1 \\
\vdots & \vdots & \vdots & \vdots & \vdots \\
UT_{m} & U6_m & U7_m & DL_m & UL_m
\end{array}\right]\left(\begin{array}{c}
c_{UT} \\
c_{U6} \\
c_{U7} \\
\alpha_d\\
\alpha_{u}
\end{array}\right)
\end{equation}

\begin{equation}
    P_j = (1 - I_{5G})\times BP_{CPU}+ I_{5G}\times BP_{5G} + UT_j\times c_{UT} + U6_j\times c_{U6} + U7_j\times c_{U7} + DL_j\times \alpha_d + UL_j\times \alpha_u
\end{equation}

$I_{5G}$ denotes the 5g indicator to indicate that whether 5g is active, if it's active $I_{5G}$ equals 1, otherwise it's 0. $P_j$ refers to total power consumption under j-th situation which is determined by all variables. $BP_{CPU}$ is the base power with no processes running. $BP_{5G}$ is the base power when the 5g connection is active but with 0 throughput. $c_{UT},c_{U6}$ and $c_{U7}$ are correspondingly the coefficients of total CPU usage(UT), CPU 6 usage(U6) and CPU 7 usage(U7). $\alpha_d$ represents the coefficients of downlink throughout and $\alpha_u$ represents the coefficients of uplink throughout.

\subsection{Thermal model}

Our thermal model is founded upon the compact thermal model framework. A well-established duality exists between the thermal and RC (resistor-capacitor) electrical networks [22]. This duality is concisely summarized in Table 2, and a graphical representation of our thermal model is provided in Figure 2. We can further streamline this model by considering it under steady-state conditions. Upon reaching steady-state temperatures, the capacitors in the circuits can be disregarded, leading to a simplified model as depicted in Figure 3.

\begin{table}[h]
\centering
\begin{tabular}{|l|l|}
\hline
Thermal            & Electrical Dual       \\ \hline
Temperature($^oC$)        & Voltage($V$)               \\
Power($W$)             & Current($A$)               \\
Thermal resistance($^oC/W$) & Electrical resistance($\Omega$) \\
Heat capacity($J/^oC$)      & Electrical capacity($F$)   \\ \hline
\end{tabular}
\caption{\label{tab:SOTA} Thermal quantities and electrical duals}
\end{table}

\begin{figure}[H]
  \centering
  \begin{minipage}[b]{0.43\textwidth}
    \includegraphics[width=\textwidth]{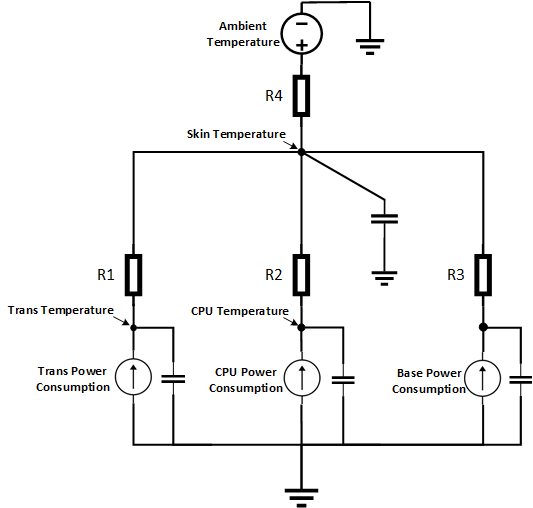}
    \caption{Thermal Model}
  \end{minipage}
  \hfill
  \begin{minipage}[b]{0.41\textwidth}
    \includegraphics[width=\textwidth]{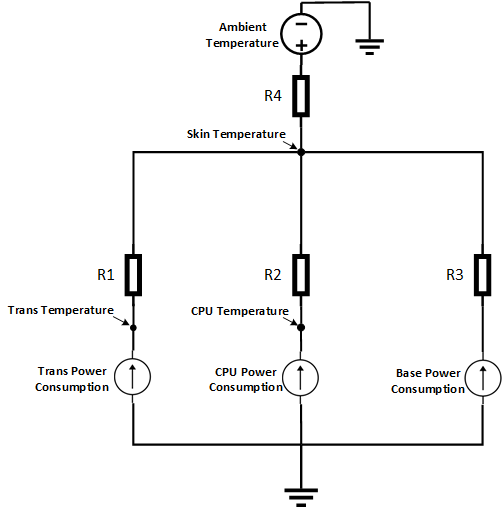}
    \caption{Steady State Thermal Model}
  \end{minipage}
\end{figure}

Then based on the circuit diagram, we can calculate the value of thermal resistor R4:
\begin{align}
    R4 &= \frac{T_{skin} - T_{amb}}{P_{CPU} + P_{Trans} + P_{base}} \\
    R4 &= \frac{T_{skin} - T_{amb}}{P_{tot}}
\end{align}

Here $T_{skin}$ denotes the skin temperature and $T_{amb}$ denotes the ambient temperature. $P_{CPU}$, $P_{Trans}$ and $P_{base}$ correspondingly refer to the power consumption of CPU, transceiver and base. $P_{tot}$ represents the total power consumption. 

As evidenced in Equation 4, the total power consumption and the temperatures of both the device skin and the ambient environment are the sole three variables required to compute R4, which can be readily captured. Subsequently, a linear regression can be utilized to obtain the estimated R4 value based on steady-state data acquired under varying ambient temperatures.

Upon determining the value of the thermal resistor R4, one can calculate the maximum acceptable power consumption under different ambient temperatures, which will not induce throttling, by merely supplying the statistics for the selected parameters.

\section{ROOT CAUSE OF SKIN TEMPERATURE RELATED THROTTLING}

In this section, we present a comparison of the total power consumption under two CPU frequency settings (High frequencies: [CPU cluster 1: 1.8 GHz, CPU cluster 2: 2.2 GHz, CPU cluster 3: 2.4 GHz]; Low frequencies: [CPU cluster 1: 1.07 GHz, CPU cluster 2: 652 MHz, CPU cluster 3: 1.4 GHz]). Each CPU frequency setting encompasses three stress cases. The first case involves stressing only the CPU to its maximum frequency and usage. In the second case, the transceiver is saturated to 2 Gbps using the Verizon mmWave service. The third case consists of stressing both the CPU and transceiver simultaneously.

\begin{figure}[!htb]\centering
   \begin{minipage}{0.49\textwidth}
     \frame{\includegraphics[width=.7\linewidth]{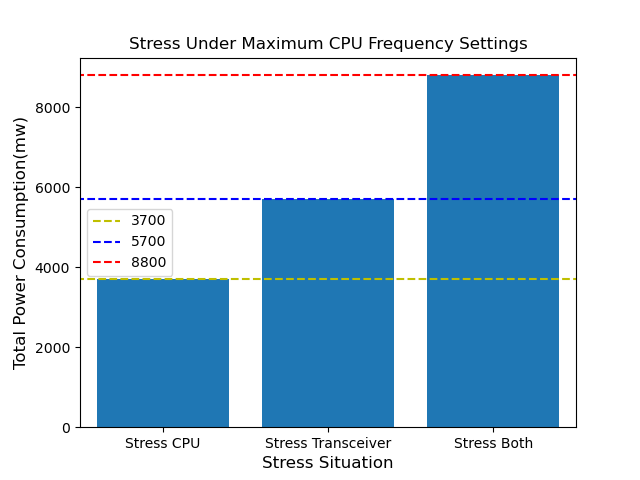}}
     \centering
     \caption{Stress Test under High CPU Frequency}\label{Fig:Data1}
   \end{minipage}
   \begin{minipage}{0.49\textwidth}
     \frame{\includegraphics[width=.7\linewidth]{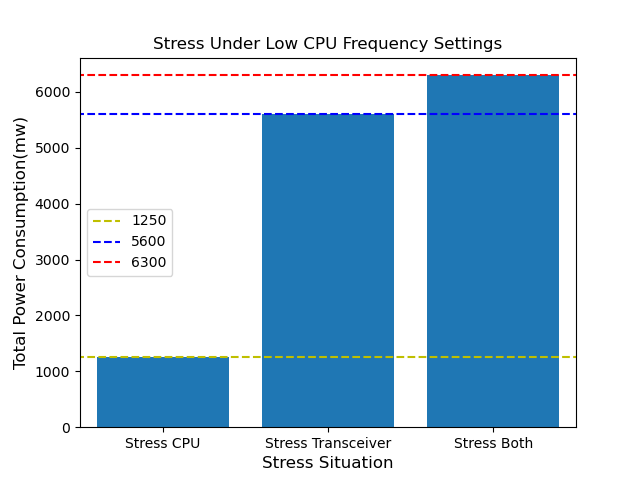}}
     \centering
     \caption{Stress Test Under Low CPU Frequency}\label{Fig:Data1}
   \end{minipage}
\end{figure}

Figure 4 reveals that in the high CPU frequency settings, when stressing the CPU and transceiver separately, the total power consumption is approximately 3700 mW and 5700 mW, respectively. When stressing both the CPU and transceiver together, the total power consumption increases to around 8800 mW. Figure 5 demonstrates that in the low CPU frequency settings, when stressing the CPU and transceiver separately, the total power consumption is approximately 1250 mW and 5600 mW, respectively. When stressing both the CPU and transceiver together, the total power consumption increases to around 6300 mW.

We observed that the sum of total power consumption in the first case (CPU only) and the second case (transceiver only) is nearly equal to the sum of total power consumption in the third case (both components) and the CPU base power (approximately 500 mW). This indicates that the CPU power consumption under the stressed transceiver case is minimal; otherwise, the power sum of the first two cases would significantly exceed the power consumption when stressing both components. Moreover, we note that the power consumption under the cases stressing only the transceiver is roughly the same across different CPU frequencies. Consequently, we can conclude that when stressing the transceiver alone, the CPU has a negligible impact on the contribution to total power consumption.

\section{POWER MODEL VALIDATION}

To evaluate the performance of the power model, we utilize three different methods RMSE, correlation coefficient $\rho$ and absolute error. The accuracy of the model can be defined using absolute error as follows:
\begin{equation}
    Accuracy = 1 - \frac{|P_e - P_m|}{P_m}\%
\end{equation}

where $P_e$ is the power estimated by the power model, and $P_m$ is the measured power. All the data given in the following sections follow the same evaluation indicator.

\begin{table}[h]
\centering
\begin{tabularx}{0.8\textwidth} 
{ 
  | >{\centering\arraybackslash}X 
  || >{\centering\arraybackslash}X 
  | >{\centering\arraybackslash}X | 
 }
 \hline
 Evaluation Metric & Training Results & Validation Results \\
 \hline
 $\rho$ &  0.978 & 0.887 \\
 \hline
 RMSE & 441.24 & 880.82 \\
 \hline
 $Accuracy$ & 0.862 & 0.834 \\
 \hline
\end{tabularx}
\caption{\label{tab:SOTA} Evaluation Results for LR Model 1}
\end{table}

We conducted hundreds of test iterations using both Verizon and T-Mobile services. Each test was performed with random transmission rates and CPU usage under specific CPU frequency settings. The average evaluation results are presented in Table 3. Two representative testing samples are depicted in Figures 6 and 7.

\begin{figure}[!htb]\centering
   \begin{minipage}{0.49\textwidth}
     \frame{\includegraphics[width=.7\linewidth]{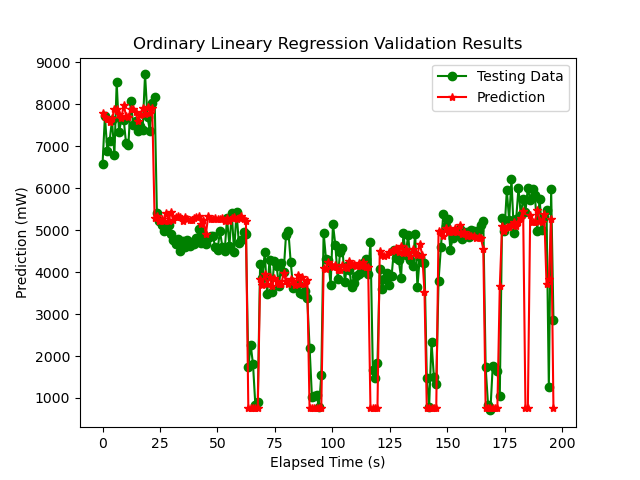}}
     \centering
     \caption{Stress Test under High CPU Frequency}\label{Fig:Data1}
   \end{minipage}
   \begin{minipage}{0.49\textwidth}
     \frame{\includegraphics[width=.7\linewidth]{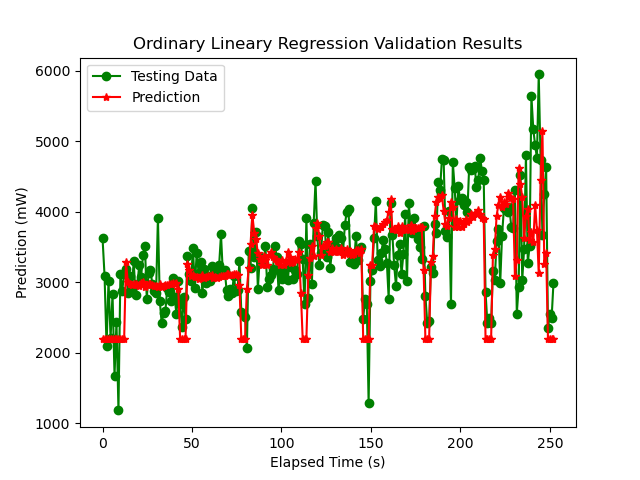}}
     \centering
     \caption{Stress Test Under Low CPU Frequency}\label{Fig:Data1}
   \end{minipage}
\end{figure}

\section{AMBIENT TEMPERATURE AWARE SUSTAINABLE THROUGHPUT}

As depicted in Figure 8, the scatter points represent the measured data under varying ambient temperatures. Upon conducting linear regression, we obtain the value of thermal resistance R4, which is 0.005$^oC/\text{mW}$. Subsequently, based on the value of R4 and the skin temperature-related throttling threshold that can be easily acquired by calling Android HardwarePropertiesManager API, we can calculate the maximum acceptable power consumption as a function of ambient temperature, as illustrated in Figure 9. Ultimately, we can determine the necessary power reduction to prevent the throttling temperature from being reached under different ambient temperatures.

\begin{figure}[H]
  \centering
  \begin{minipage}[b]{0.48\textwidth}
    \includegraphics[width=\textwidth]{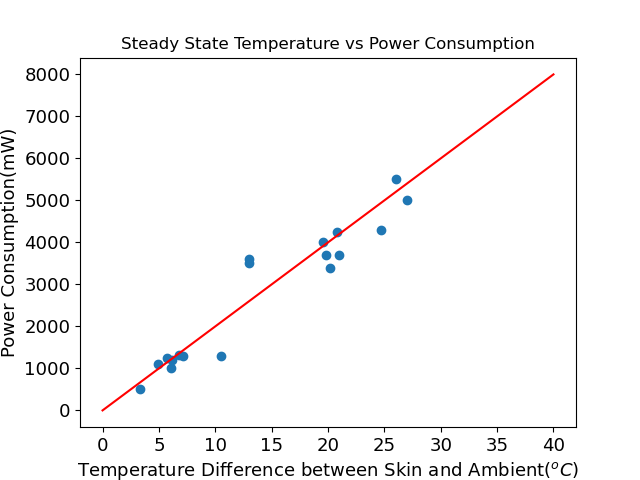}
    \caption{Linear Regression of the Data Captured under Different Ambient Temperature}
  \end{minipage}
  \hfill
  \begin{minipage}[b]{0.43\textwidth}
    \includegraphics[width=\textwidth]{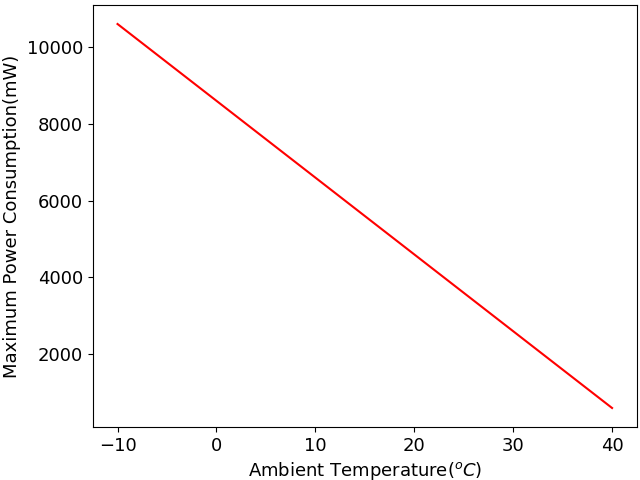}
    \caption{Maximum Acceptable Power Consumption as a Function of Ambient Temperature}
  \end{minipage}
\end{figure}

\begin{figure}[H]
  \centering
  \begin{minipage}[b]{0.48\textwidth}
    \includegraphics[width=\textwidth]{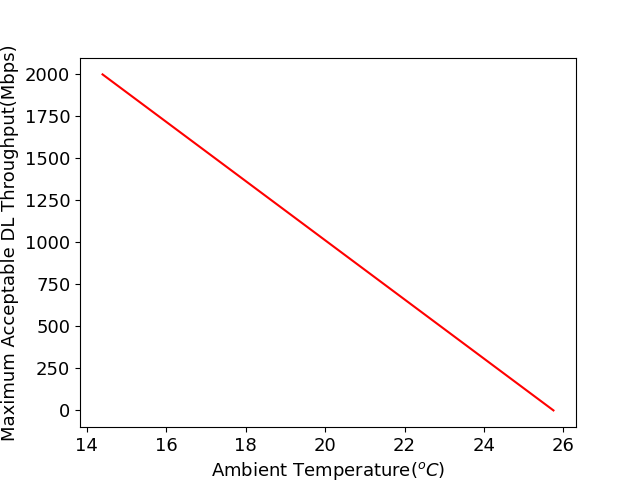}
    \caption{Maximum Data Rate without throttling as a Function of Ambient Temperature}
  \end{minipage}
  \hfill
  \begin{minipage}[b]{0.43\textwidth}
    \includegraphics[width=\textwidth]{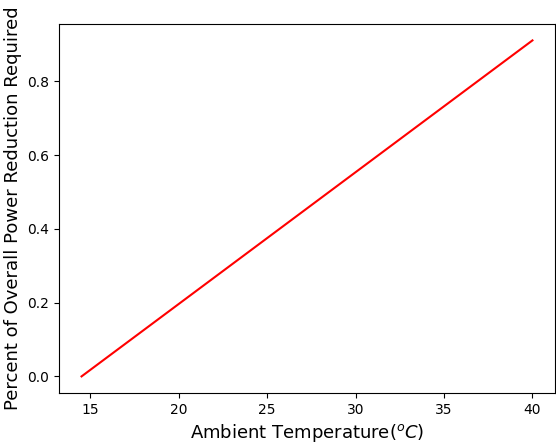}
    \caption{Percent Overall Power Reduction Required as a Function of Ambient Temperature}
  \end{minipage}
\end{figure}

Figure 10 displays the relationship between ambient temperature and the maximum data rate achievable without throttling. From our previous experiments, we know that the overall power consumption with the maximum 2 Gbps downlink throughput and 0 uplink throughput is approximately 5700 mW. Consequently, we regard 5700 mW as the power consumption with the highest throughput. This allows us to establish the relationship between the percentage of overall power reduction required to prevent throttling and ambient temperature, as shown in Figure 11.

\section{CONCLUSION}

This paper concentrates on examining skin temperature-related throttling in 5G mmWave communication. We have developed a power and thermal model for 5G mmWave smartphones, which is subsequently employed to ascertain that the transceiver is the root cause of throttling. Furthermore, we investigate the maximum non-throttling transmission rate and the power reduction necessary to prevent throttling under specific ambient temperatures. The methodology proposed in this paper can be seamlessly extended to the majority of commercially available mmWave smartphones with skin temperature sensors. Our research holds potential guiding significance for future optimization endeavors, encompassing both software and hardware aspects.

\section{FUTURE WORK}
The optimization of mmWave technology remains an open research topic, garnering significant attention from researchers. Efforts to address overheating problems in mmWave devices primarily focus on hardware and software solutions. From a software perspective, the primary strategies include Radio Resource Management Optimization and Energy-efficient Compression Technologies. On the hardware side, the primary focus is RF Antenna Circuit Optimization. 

We are currently evaluating existing compression technologies employed in 5G mmWave systems to explore potential improvements. Additionally, our efforts are underway to design optimized mmWave RF circuits, particularly concentrating on the Low Noise Amplifier (LNA) component. We are trying to replace the traditional power source in the LNA circuit with a voltage source, which can potentially reduce power consumption caused by the biasing circuit. This method has the potential to decrease the overall power consumption for 5G mmWave transmissions and eventually increase the maximum sustainable throughput without causing throttling.

~\\

\noindent \textbf{\Large References}

[1] Rochman, Muhammad Iqbal, Damian Fernandez, Norlen Nunez, Vanlin Sathya, Ahmed S. Ibrahim, Monisha Ghosh, and William Payne. "Impact of Device Thermal Performance on 5G mmWave Communication Systems." arXiv preprint arXiv:2202.04830 (2022).

[2] NR; Overall Description; Stage-2, document TS 38.300, 3GPP,
Version 16.1.0, Apr. 2020

[3] Saadat, Moh Sabbir, Sanjib Sur, and Srihari Nelakuditi. "Bringing temperature-awareness to millimeter-wave networks." In Proceedings of the 26th Annual International Conference on Mobile Computing and Networking, pp. 1-3. 2020.

[4] Huang, Junxian, Feng Qian, Alexandre Gerber, Z. Morley Mao, Subhabrata Sen, and Oliver Spatscheck. "A close examination of performance and power characteristics of 4G LTE networks." In Proceedings of the 10th international conference on Mobile systems, applications, and services, pp. 225-238. 2012.

[5] Narayanan, Arvind, Xumiao Zhang, Ruiyang Zhu, Ahmad Hassan, Shuowei Jin, Xiao Zhu, Xiaoxuan Zhang et al. "A variegated look at 5G in the wild: performance, power, and QoE implications." In Proceedings of the 2021 ACM SIGCOMM 2021 Conference, pp. 610-625. 2021.

[6] Yuan, Xinjie, Mingzhou Wu, Zhi Wang, Yifei Zhu, Ming Ma, Junjian Guo, Zhi-Li Zhang, and Wenwu Zhu. "Understanding 5G performance for real-world services: a content provider's perspective." In Proceedings of the ACM SIGCOMM 2022 Conference, pp. 101-113. 2022.

[7] T. Ma, M. Hempel, D. Peng and H. Sharif, "A Survey of Energy-Efficient Compression and Communication Techniques for Multimedia in Resource Constrained Systems," in IEEE Communications Surveys $\&$ Tutorials, vol. 15, no. 3, pp. 963-972, Third Quarter 2013, doi: 10.1109/SURV.2012.060912.00149.

[8] Al-Kadhim, Halah Mohammed, and Hamed S. Al-Raweshidy. "Energy efficient data compression in cloud based IoT." IEEE Sensors Journal 21, no. 10 (2021): 12212-12219.

[9] A. Huang, K. -H. Lin and H. -Y. Wei, "Thermal Performance Enhancement With DRX in 5G Millimeter Wave Communication System," in IEEE Access, vol. 9, pp. 34692-34707, 2021, doi: 10.1109/ACCESS.2021.3061728.

[10] M. Lauridsen, D. Laselva, F. Frederiksen and J. Kaikkonen, "5G New Radio User Equipment Power Modeling and Potential Energy Savings," 2019 IEEE 90th Vehicular Technology Conference (VTC2019-Fall), 2019, pp. 1-6, doi: 10.1109/VTCFall.2019.8891215.

[11] T. Kim et al., "Evolution of Power Saving Technologies for 5G New Radio," in IEEE Access, vol. 8, pp. 198912-198924, 2020, doi: 10.1109/ACCESS.2020.3035186.

[12] Wang, Hua, Song Hu, Taiyun Chi, Fei Wang, Sensen Li, Min-Yu Huang, and Jong Seok Park. "Towards energy-efficient 5G mm-wave links: Exploiting broadband mm-wave Doherty power amplifier and multi-feed antenna with direct on-antenna power combining." In 2017 IEEE Bipolar/BiCMOS Circuits and Technology Meeting (BCTM), pp. 30-37. IEEE, 2017.

[13] Alaji, Issa, Etienne Okada, Daniel Gloria, Guillaume Ducournau, and Christophe Gaquière. "Temperature compensated power detector towards power consumption optimization in 5G devices." Microelectronics Journal 120 (2022): 105351.

[14] Bai, Qing, and Josef A. Nossek. "Energy efficiency maximization for 5G multi‐antenna receivers." Transactions on Emerging Telecommunications Technologies 26, no. 1 (2015): 3-14.

[15] NR and NR-RAN Overall Description; Stage 2 (Release 15), V15.4.0,
document TS 38, 3GPP, Dec. 2018. [Online]. Available: http://ftp.3gpp.org.

[16]  I. L. Da Silva, G. Mildh, M. Saily, and S. Hailu, ‘‘A novel state model for
5G radio access networks,’’ in Proc. IEEE Int. Conf. Commun. Workshops
(ICC), Kuala Lumpur, Malaysia, May 2016, pp. 632–637.

[17] Giordani, Marco, and Michele Zorzi. "Satellite communication at millimeter waves: A key enabler of the 6G era." In 2020 International Conference on Computing, Networking and Communications (ICNC), pp. 383-388. IEEE, 2020.

[18] Wang, Xiong, Linghe Kong, Fanxin Kong, Fudong Qiu, Mingyu Xia, Shlomi Arnon, and Guihai Chen. "Millimeter wave communication: A comprehensive survey." IEEE Communications Surveys $\&$ Tutorials 20, no. 3 (2018): 1616-1653.

[19] Qiao, Jian, Yejun He, and Xuemin Sherman Shen. "Proactive caching for mobile video streaming in millimeter wave 5G networks." IEEE Transactions on Wireless Communications 15, no. 10 (2016): 7187-7198.

[20] Sukhmani, Sukhmani, Mohammad Sadeghi, Melike Erol-Kantarci, and Abdulmotaleb El Saddik. "Edge caching and computing in 5G for mobile AR/VR and tactile internet." IEEE MultiMedia 26, no. 1 (2018): 21-30.

[21] Tokoro, Setsuo. "Automotive application systems of a millimeter-wave radar." In Proceedings of Conference on Intelligent Vehicles, pp. 260-265. IEEE, 1996.

[22] M. Pedram and S. Nazarian, "Thermal Modeling, Analysis, and Management in VLSI Circuits: Principles and Methods," in Proceedings of the IEEE, vol. 94, no. 8, pp. 1487-1501, Aug. 2006, doi: 10.1109/JPROC.2006.879797.

\end{document}